\documentclass[twocolumn]{aa}

\usepackage{graphicx}
\usepackage{txfonts}
\begin{document}

   \title{NICER observations of the Crab pulsar glitch of 2017 November}

   \author{M. Vivekanand \inst{1}}

   \institute{No. 24, NTI Layout 1\textsuperscript{st} Stage, 
              3\textsuperscript{rd} Main, 1\textsuperscript{st} Cross, 
	      Nagasettyhalli, Bangalore 560094, India. \\
   \email{viv.maddali@gmail.com}}

   \date{}

 
  \abstract
   {
   The Crab pulsar underwent its largest timing glitch on 2017 Nov 8.
   The event was discovered at radio wavelengths, and was followed at 
   soft X-ray energies by observatories, such as XPNAV and NICER.
   }
   {
   This work aims to compare the glitch behavior at the two 
   wavelengths mentioned above. Preliminary work in this regard has been done by the 
   X-ray satellite XPNAV. NICER with its far superior sensitivity is 
   expected to reveal much more detailed behavior.
   }
   {
   NICER has accumulated more than $301$ kilo seconds of data on 
   the Crab pulsar, equivalent to more than $3.3$ billion soft X-ray photons.
   These data were first processed using the standard NICER analysis 
   pipeline. Then the arrival times of the X-ray photons were 
   referred to the solar system's barycenter. Then specific analysis
   was done to study the specific behavior outlined in the following 
   sections, while taking dead time into account.
   }
   {
   The variation of the rotation frequency of the Crab pulsar and 
   its time derivative during the glitch is almost exactly similar
   at the radio and X-ray energies. The following properties of the Crab 
   pulsar remain essentially constant before and after the glitch: the 
   total X-ray flux; the flux, widths, and peaks of the two components of 
   its integrated profile; and the soft X-ray spectrum. There is
   no evidence for giant pulses at X-ray energies. However, the timing noise 
   of the Crab pulsar shows quasi sinusoidal variation before the glitch, 
   with increasing amplitude, which is absent after the glitch.
   }
   {
   Even the strongest glitch in the Crab pulsar appears not to affect 
   all but one of the properties mentioned above, at either frequency. 
   The fact that the timing noise appears to change due to the glitch 
   is an important clue to unravel as this is still an unexplained phenomenon.
   }


   \maketitle
%

\section{Introduction}

On $2017$ Nov $8$, a timing glitch occurred in the Crab pulsar that was the largest 
of the glitches so far \citep{Shaw2018}. They analyzed data for about $150$ days after the 
glitch, and several hundred days before, and brought about the following conclusions that 
are relevant for this work.

Firstly, the rotation frequency $\nu$ rises abruptly at the glitch and decays 
exponentially, which is similar to other glitches of the Crab pulsar. Most of this rise is abrupt
(unresolved in time) while a small part of the rise is delayed and resolved
in their data (panel C of Fig.~$1$, and top panel of Fig.~$3$, of \cite{Shaw2018}).
Next, the time derivative of the frequency $d \nu / d t = \dot \nu$ reflects the variation 
of $\nu$ on short time scales ($< 10$ days) -- unresolved increase and rapid 
decrease. On longer timescales, it exponentially recovers to values close to the 
preglitch value, similar to other glitches of the Crab pulsar (panel D of Fig.~$1$,
and bottom panel of Fig.~$3$, of \cite{Shaw2018}).
Further, there is no change in pulse morphology due to the glitch at radio frequencies. 
The parameters that remain constant across the glitch are the peaks and widths of the main
pulse and the inter pulse of the Crab pulsar  at $610$ and $1520$ mega Hertz (MHz) 
(Fig.~$5$ of \cite{Shaw2018}).
Finally, there is no change in the $2 - 50$ kilo electron volt (Kev) X-ray flux on 
account of the glitch (Fig.~$6$ of \cite{Shaw2018}).

Some of these results were confirmed by the X-Ray Pulsar Navigation (XPNAV) 
satellite \citep{Zhang2018},
whose main purpose was navigation and not astronomy. It nevertheless obtained 
unprecedented cadence of timing observations at X-ray energies in the range
$0.5 - 10$ Kev. See their Fig.~$3$ for the variation in $\nu$ and $\dot \nu$ at the
glitch, and their Fig.~$5$ for constancy of X-ray flux across the glitch.

This work analyses the X-ray data obtained by the Neutron star Interior Composition 
Explorer (NICER) satellite \citep{Gendreau2017},
that is significantly more sensitive than XPNAV, which allows for some
analysis that is not possible with XPNAV data. However, NICER is not a dedicated timing
instrument, so its cadence of timing observations is nowhere near that of XPNAV.

\section{Observations and analysis}

NICER consists of $56$ coaligned detectors, each of which is essentially a small X-ray 
telescope \citep{Arzoumanian2014, Gendreau2016, Prigozhin2016}. The operating range of NICER is 
$0.2 - 12$ Kev with a peak collecting area of $1900$ cm$^{2}$ at $1.5$ Kev, while that 
of XPNAV is $2.4$ cm$^{2}$ at $1.5$ Kev. The time resolution is better than a microsecond 
($\mu$s); photons are time stamped by an onboard GPS.

At the time of carrying out this work, NICER observed the Crab pulsar on $62$ different 
days, starting from $2017$ Aug $5$ 
to $2019$ Apr $26$, which is a duration of $630$ days. The first observation was
$94$ days before the glitch (observation identity number (ObsID) $1013010101$, 
Modified Julian Date (MJD) $57970.791$). There were eight days of observations during
$2017$ Aug, and two days in $2017$ Sep, the last of which occurred on $2017$ Sep $28$
(ObsID $1011010201$, MJD $58024.575$). Then there was a gap of $42$ days before
the next observation (ObsID $1013010109$, MJD $58066.421$), which was two days after 
the glitch (glitch epoch MJD $58064.555$), but this observation had a live time
of only $224$ seconds (sec). For the next $11$ days, the observations were done daily,
after which there was a gap of $40$ days (ObsID $1013010122$, MJD $58117.329$).
From then on the observations were mostly on individual days separated by long gaps, 
except for four continuous days of observation in $2018$ Mar, eight continuous days of 
observation in $2018$ Sep, and four continuous days of observation in $2018$ Dec.
The last observation available for this work was on $2019$ Apr $26$ (ObsID $2013010103$, 
MJD $58599.988$), but this had only $34$ sec of live time. This kind of 
nonuniform cadence of observations implies that one can not perform the phase coherent 
timing analysis that was done by \cite{Shaw2018} and \cite{Zhang2018}. This is 
particularly true for the glitch that is under discussion; it is so large that the change 
in phase over a single day can be larger than one cycle so that the cadence required
for phase coherent analysis is several timing observations in a single day.

Five of these ObsIDs have live times of $2$, $7$, $20$, $34$, and $92$ sec, 
and were of no use since one requires at least $100$ sec of live time on the 
Crab pulsar to obtain a reasonable integrated profile (IP). In addition, two pairs
of observations (ObsIDs $1013010131$ and $1013020101$ observed on $2018$ Apr $7$,
and $2013010101$ and $2013020101$ observed on $2019$ Mar $11$) appear to have 
been observed in some experimental mode -- their photon events are not time
ordered, but distributed across the two files. So they could not be analyzed
together. For Section $4$, which required the alignment of the data of individual 
ObsIDs, only those having live times greater than $1000$ sec were used. This 
yielded $43$ useful ObsIds for analysis. For Section $3$ which analyses several
ObsIDs together, smaller files with at least $100$ sec of live time could also 
be used.

The data were analyzed using NICER version five software included in the HEAsoft 
distribution 6.25; the calibration setup was the CALDB version XTI(20190516). The analysis
began with the pipeline tool {\it{nicerl2}} with default parameters, which selects
all $56$ detectors, applies standard filters, applies calibration, cleans the events
and merges them, etc. Next the average count rate for each detector was calculated using
the {\it{extractor}} tool in the light curve mode, and those with counts rates 
significantly below the mean value are excluded; see \cite{Deneva2019} for 
details. In the list of $43$ ObsIds of Section $4$, $40$ had $52$ useful detectors, 
while two had $51$ detectors and one had $50$ detectors.  Next, light curves were 
scrutinized for count rates well above and below the mean value. These segments of 
data were excluded by editing the good times in the tenth HDU (GTI\_FILT), and then 
using the tool {\it{fselect}} with the {\it{gtifilter}} option to filter out photon 
events outside the good times. The event epochs were then referenced to the solar 
system barycenter using the {\it{barycorr}} tool using the JPL ephemeris DE$431$, 
with the position of the Crab pulsar at that epoch as input.

Finally, an approximate period ($1 / \nu$) is estimated (at a nominal period 
derivative ($- \dot \nu / \nu^2$)) for each ObsID by first obtaining the power 
spectrum of the data using the {\it{powspec}} tool. This is refined by searching 
for maximum probability of period using the {\it{efsearch}} tool. This is further 
refined by cross correlating the first and second halves of the data (see 
\cite{Vivekanand2015} for details). These intermediate period and period 
derivative (or alternately the intermediate $\nu$ and $\dot \nu$) are the 
starting point for the analysis of the following sections.

\subsection{Dead time}

\begin{figure}[h]
\centering
\includegraphics[width=8.5cm]{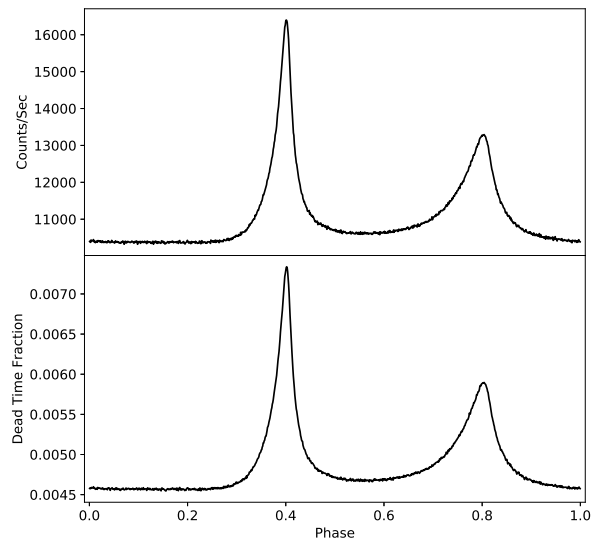}
\caption{
	Dead time estimation for the data of ObsID $1013010147$. The top panel 
	displays the IP of the Crab pulsar using the 
	appropriate values of $\nu$ and $\dot \nu$ for this ObsID (see Section 
	$3$), over $1024$ phase bins. The bottom panel displays the dead time, 
	as a fraction of live time, in the corresponding phase bins.
        }
\label{fig1}
\end{figure}

An important advantage of NICER's design is that the effective dead time is very low 
\citep{LaMarr2016, Stevens2018}. However this has not been estimated quantitatively
so far for the Crab pulsar. Fig.~\ref{fig1} shows the dead time for the longest 
observation in the data set.

Typically, each photon event of NICER results in a dead time of about $15$ or $22$ 
$\mu$s for each detector \citep{LaMarr2016}; practically the dead time is larger. 
Now the typical photon count rate per detector, for the Crab nebula plus 
the Crab pulsar, can be estimated by dividing the total number of photons obtained
in Fig.~\ref{fig1} by the live time and by the number of detectors, which is
$260699397 / 23722.5 / 52 \approx 211.3$ counts per second per detector. This 
implies a mean interval of about $4.7$ milliseconds (ms) between photons. Clearly 
the dead time is a negligible fraction of this interval, so the dead time correction 
to the observed count rate would also be negligible. Because the dead time is 
such a small value, it scales almost linearly with the photon count rate in each phase 
bin of the IP in Fig.~\ref{fig1}, as expected. The offpulse 
dead time, in the phase range $0 - 0.2$, is $0.46$ per cent (\%) of the live time; 
at the peak of the Crab pulsar's pulse the dead time fraction is $0.73$\%. 

These numbers are consistent with the expected values. A Poisson process with mean
interval between events $\tau$ has an exponential probability density 
distribution for the time interval between any two events: $1/\tau \times \exp 
\left ( -t / \tau \right )$. So the fraction of events that fall during the dead 
time, say $T$ sec, is (by simple integration) $\approx T / \tau$. Now, the mean 
dead time for the above ObsID is $T = 23.1 \pm 4.2$ $\mu$s, and $\tau = 4.7$ ms, 
so the mean dead time fraction is $\approx 0.49_{-0.09}^{+0.09}$\%.
This work estimates the dead time as a function of the phase of the Crab pulsar's 
pulse for each ObsID, and uses it for correction throughout this work, although 
it is indeed a negligible value. 

\section{Stride fit}

\begin{figure}[h]
\centering
\includegraphics[width=8.5cm]{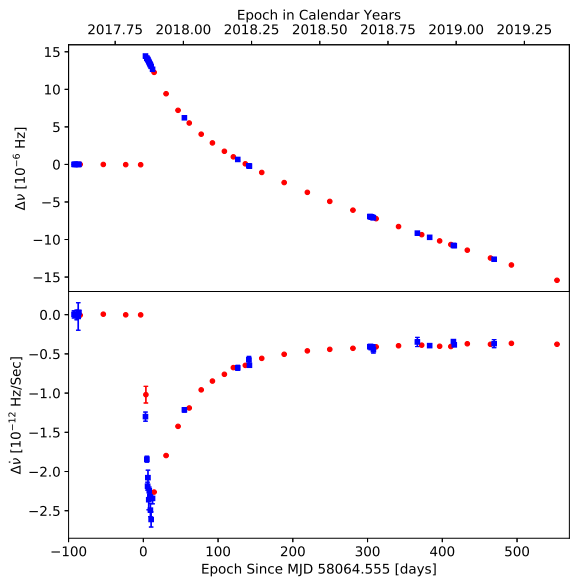}
\caption{
	Variation of $\nu$ and $\dot \nu$ as a function of epoch.
	Dots are data from the JBCPME \citep{Lyne1993}, while the boxes are NICER 
	data. The change in rotation frequency $\Delta \nu$ (top panel) and the 
	change in the frequency derivative $\Delta \dot \nu$ (bottom panel) are 
	measured with respect to the preglitch reference timing model obtained 
	from the preglitch JBCPME data.
        }
\label{fig2}
\end{figure}

The top and bottom panels of Fig.~$3$ of \cite{Shaw2018} display the long term 
variation of $\nu$ and $\dot \nu$, respectively, at radio wavelengths. The top
and bottom panels of Fig.~$3$ of \cite{Zhang2018} do the same at soft X-ray
energies. \cite{Shaw2018} obtain their results by what they label as the 
``striding boxcar'' fit, which essentially fits their timing residuals in 
smaller but overlapping segments of epochs. Thus they fit timing residuals for 
$20$ consecutive days to a second order polynomial to obtain the $\nu$ and 
$\dot \nu$ at the central epoch of the data. They repeat the exercise after 
sliding the $20$ day ``boxcar'' by five days, to obtain the $\nu$ and $\dot \nu$ 
at the next epoch; this implies a $15$ day overlap of data between adjacent
boxcars. This is done after averaging their original data to just two timing
residuals per day, for higher sensitivity to explore the long term variation.
\cite{Zhang2018} do the same stride fit, but their boxcar is four days long, 
with sliding step of half a day.

Fig.~\ref{fig2} shows the results of applying the stride fit technique to NICER
data of the Crab pulsar. Since the data cadence is highly nonuniform and
inadequate, only that data is analyzed in this section that consists of at least
two days of consecutive observations. The boxcar width used is three days, except 
when only two days are possible; the sliding step is one day. Timing residuals are 
estimated for each $100$ sec of data within the boxcar, provided the actual
live time within the $100$ sec is at least $90$ sec. Further details of
the stride fit analysis are given in Appendix A.

\begin{table}
\begin{center}
\caption{The preglitch reference timing model, obtained using $\nu$ from JBCPME 
	\citep{Lyne1993} for $\approx 700$ days before the glitch. The errors ($1 \sigma$)
in the last digit of each number are shown in brackets. \label{tbl1}
}
\begin{tabular}{|l|c|}
\hline
Parameter  & Value \\
\hline
Glitch Epoch (MJD) & $58064.555$ \\
\hline
$\nu_0$ (Hz) & $29.636716899(9)$ \\
\hline
$\dot \nu_0$ ($10^{-10}$ Hz s$^{-1}$) & $-3.68604(1)$ \\
\hline
$\ddot \nu_0$ ($10^{-20}$ Hz s$^{-2}$) & $1.2(1)$ \\
\hline
\hline
\end{tabular}
\end{center}
\end{table}

The dots in Fig.~\ref{fig2} are the departures of $\nu$ and $\dot \nu$ values 
tabulated in the so called Jodrell Bank Crab Pulsar Monthly 
Ephemeris\footnote{http://www.jb.man.ac.uk/pulsar/crab.html} (\cite{Lyne1993}; 
henceforth JBCPME), with respect to the preglitch timing model. This is
obtained by fitting the frequency and its first two time derivatives, at the 
glitch epoch, to data $700$ days prior to the glitch; these values ($\nu_0$, 
$\dot \nu_0$ and $\ddot \nu_0$) are given in Table~\ref{tbl1}. The departures 
$\Delta \nu$ and $\Delta$ $\dot \nu$ vary as in Fig.~$3$ of \cite{Shaw2018}.

The values of $\nu_0$, $\dot \nu_0$ and $\ddot \nu_0$ of \cite{Shaw2018} are 
given in their Table~$1$. Propagating them from their reference epoch to the 
glitch epoch, their $\nu_0$ differs from the value in Table~\ref{tbl1} above 
by $-0.492$ micro Hertz ($\mu$Hz); their $\dot \nu_0$ differs from the above 
value by $0.042 \times 10^{-12}$ Hz/sec. These two values are negligible 
compared to the scale of $\Delta \nu$ and $\Delta \dot \nu$ in Fig.~\ref{fig2}, 
respectively. Strictly, however, they are significant compared to their formal 
errors. The $\ddot \nu_0$ of \cite{Shaw2018} is an order of magnitude larger 
than the value in Table~\ref{tbl1}. 

Undertaking a similar exercise for the preglitch parameters of \cite{Zhang2018},
their $\nu_0$ differs from the value in Table~\ref{tbl1} above by $-0.490$ 
$\mu$Hz; their $\dot \nu_0$ differs from the above value by $-0.005 \times 
10^{-12}$ Hz/sec. The $\ddot \nu_0$ of \cite{Zhang2018} is an factor of two 
smaller than the value in Table~\ref{tbl1}. 

I believe these differences arise from the fact that firstly, the $\nu$ values 
listed by the JBCPME (from which Table~\ref{tbl1} is derived) are themselves 
average values, mostly monthly averages, and secondly the preglitch durations 
of the three works may differ significantly; this may particularly affect the 
frequency second derivative. It is therefore concluded that there is broad 
agreement between the preglitch parameters derived here with those of 
\cite{Shaw2018} and \cite{Zhang2018}, at least for the purpose of this work.

The boxes in Fig.~\ref{fig2} are the departures $\Delta \nu$ and $\Delta \dot 
\nu$ estimated from NICER data, with respect to the preglitch timing model given 
in Table~\ref{tbl1}. The agreement between the dots (radio data) and the boxes
(soft X-ray NICER data) is excellent, not just very close to the glitch but
also almost $500$ days away from it. The XPNAV data covered only $100$
days after the glitch; within these $100$ days, Fig.~\ref{fig2} above is 
both qualitatively and quantitatively consistent with Fig.~$3$ of
\cite{Zhang2018}.
This section therefore concludes that the variation of $\nu$ and $\dot \nu$
of the Crab pulsar during the glitch of $2017$ Nov $8$ is very similar at
the radio and X-ray energies, right from $\approx 100$ days before the glitch
to almost $500$ days afterward.

\section{Pulse properties}

At $610$ MHz radio frequency, the IP of the 
Crab pulsar consists of a main pulse; an inter pulse about $0.4$ phase 
cycles after the main pulse, of about half the amplitude; and a much 
smaller third component known as the precursor leading the main pulse by 
$0.04$ phase cycles \citep{Shaw2018}. All three components are very narrow 
compared to one phase cycle. At $1520$ MHz the precursor almost disappears. 
\cite{Shaw2018} investigate the possibility of these components changing 
due to the glitch. They model the two main components using gaussians, and 
plot their widths, and ratio of their peaks, at both radio frequencies, 
as a function of epoch at a cadence of once a day (see their Fig.~$5$). 
They find no evidence of any change in these parameters on account of 
the glitch.

\cite{Shaw2018} also plot the daily X-ray flux of the Crab pulsar, in the 
energy range $15 - 50$ Kev, from the Burst Alert Telescope (BAT) instrument 
aboard the Swift X-ray satellite, as well as the daily X-ray flux in the 
energy range $2 - 20$ Kev from the Monitor of All-sky X-ray Image (MAXI) 
instrument aboard the International Space Station (ISS); see their Fig.~$6$. 
They conclude that there is no change in the X-ray flux of
the Crab pulsar that can be associated with the glitch.
\cite{Zhang2018} plot five day averages of the onpulse flux in the
energy range $0.5 - 10$ Kev, and notice no significant change in X-ray
flux of the Crab pulsar due to the glitch; see their Fig.~$5$.

\begin{figure}[h]
\centering
\includegraphics[width=8.5cm]{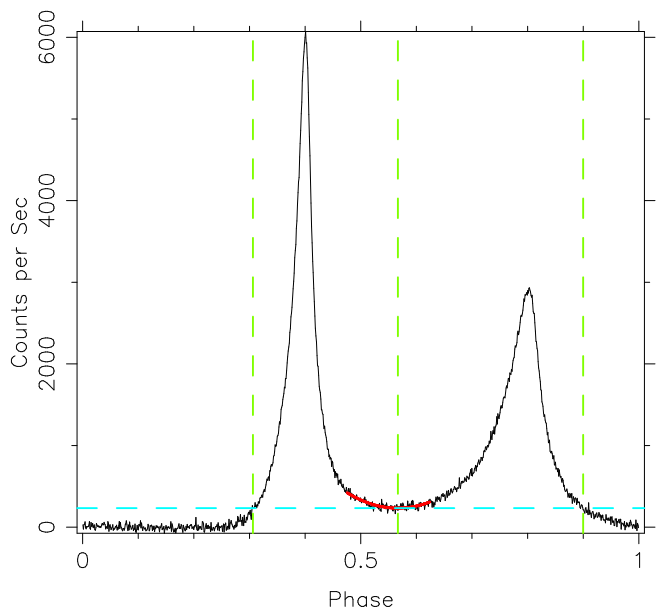}
\caption{
	Integrated profile of the Crab pulsar for the data of ObsID 
	$1013010143$. The solid red curve between the two pulse components,
	and the horizontal and vertical dashed lines define the two 
	pulse components for the purpose of parameter estimation; 
	they are explained in the text.
        }
\label{fig3}
\end{figure}

Fig.~\ref{fig3} shows the IP of the Crab pulsar for the
data of ObsID $1013010143$, after folding at the $\nu$ and $\dot \nu$
derived in the previous section, and after aligning it with the IP
in Fig.~\ref{fig1} using cross-correlation, and after subtracting the offpulse flux as 
described later. There are only two pulse components at soft X-rays, and 
both are fairly wide and appear merged into each other.  For the purpose 
of this section, the two pulse components are defined as follows. First, 
the minimum of the X-ray flux between the two peaks is identified, by 
fitting a second order polynomial to the flux data in the phase range 
$0.475 - 0.625$. This is the solid red curve between the two pulse components 
in Fig.~\ref{fig3}. Differentiating this curve gives the phase of minimum 
flux, which is identified by the middle vertical dashed line in 
Fig.~\ref{fig3}; the horizontal dashed line represents the minimum flux 
at this phase.

Next the two outer intersection points of the horizontal dashed line 
with the IP are estimated, after passing the IP data through a moving 
average filter of three phase bins, to reduce noise; these are the 
first and third vertical dashed lines in Fig.~\ref{fig3}. Then the 
main and inter pulses of the Crab pulsar's IP are identified as lying 
between the first two and last two vertical dashed lines, respectively. 
The onpulse flux of the Crab pulsar is defined as lying beyond the phase 
$0.2$.

This procedure is applied to data of each ObsID that has live time
$> 1000$ sec, after aligning (using cross-correlation) its IP with the 
reference IP in Fig.~\ref{fig1}, viz. that of ObsID $1013010147$,
and after subtracting the average offpulse flux in phase range $0.0 - 
0.2$. Then the following seven parameters are estimated: the average 
fluxes of the main pulse, the inter pulse and the onpulse; and the rms 
widths and peaks of the main and inter pulse.

\begin{figure}[h]
\centering
\includegraphics[width=8.5cm]{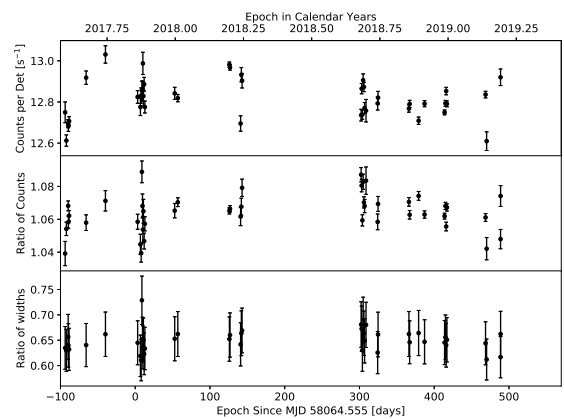}
\caption{
	Variation of pulse properties as a function of epoch.
	Top to bottom panels: Average onpulse X-ray flux of the Crab pulsar
	(photons per second per detector in the energy range $0.2 - 12$ Kev), 
	ratio of the average fluxes of the main and inter pulse, and ratio of 
	the rms widths of the main and inter pulse, respectively.
        }
\label{fig4}
\end{figure}

Fig.~\ref{fig4} shows the results of this section. The top panel
displays the average onpulse X-ray flux of the Crab pulsar per
detector. The variation of the data is consistent with it being 
essentially constant. The total range of data is $13.03 - 12.61 = 0.42$ 
photons per second per detector, while the rms of these data values
is 0.09; so the data is spread over 0.42/0.09 = 4.7 standard deviations. 
Now, for $43$ data points one expects a spread over at least three standard 
deviations. The rest is probably due to factors such as time variability 
of the calibration of the $56$ individual detectors, day and night differences 
in calibration of NICER, etc. Although there are only six of the $43$ 
points before the glitch, the variation of the data is similar 
before and after the glitch. It is therefore clear that there is no 
significant variation of this parameter due to the glitch. This is 
consistent with the results of \cite{Shaw2018} and \cite{Zhang2018}.

The middle panel of Fig.~\ref{fig4} shows the ratio of the average 
fluxes of the main and inter pulse. This is close to the value one 
because the inter pulse is wider although its peak is smaller. The
bottom panel of Fig.~\ref{fig4} shows the ratio of the rms widths
of the main and inter pulse. These two parameters also do not display
any significant variation due to the glitch. This is also consistent
with the results of \cite{Shaw2018}.
Since the lower two panels of Fig.~\ref{fig4} imply that (consequently)
the ratio of the peaks of the two components would also be constant, 
this result has not been plotted in Fig.~\ref{fig4}.

\section{Soft X-ray spectrum}

This section investigates whether the soft X-ray spectrum of the Crab pulsar
changes during the glitch. The calibrated spectrum has to be obtained
using the tool {\it{xspec}}. However, it is known that these are early
days for the NICER project, and their spectrum calibration is still
preliminary. See Section $2.1$ and Fig.~$1$ of \cite{Ludlum2018}, and 
Section $3$ of \cite{Miller2018}; see also the pdf file available in 
{\it{nicer\_arfrmf\_20180329.tar.gz}} at the NICER archive 
site\footnote{https://heasarc.gsfc.nasa.gov/docs/nicer/archive/}, and
{\it{NICERDAS-CalibGuide-20180814.pdf}} at the NICER calibration
site\footnote{https://heasarc.gsfc.nasa.gov/docs/heasarc/caldb/nicer/docs/xti/}.
Even though it is possible that several calibrations issues may have been
resolved in the later calibration files that are used in this work, the analysis 
of this section is performed on both the calibrated as well as the raw 
spectrum of the Crab pulsar.

One begins the analysis by repeating the initial procedure of Section $4$ --
data of each ObsID is folded at the appropriate $\nu$ and $\dot \nu$, and 
shifted in phase using cross-correlation so that its IP is aligned to the reference
profile in Fig.~\ref{fig1}. Then the phase of each photon is written into
an additional column in the event file. This file is filtered in phase using
the tool {\it{fselect}} to produce two event files -- one with phase $\le 0.2$ 
and another with phase $> 0.2$, corresponding to offpulse and onpulse, 
respectively. These event files are input to the tool {\it{extractor}} used 
in the spectrum mode, to produce the corresponding offpulse and onpulse raw 
spectra. To use them in {\it{xspec}}, the live times of these files have
to be corrected -- the ``EXPOSURE'' keyword in the off and on spectra is
multiplied by the factor $0.2$ and $0.8$, respectively, and inserted
using the tool {\it{fparkey}}.

Next, the keywords ``RESPFILE'' and ``ANCRFILE'' in the onpulse spectrum
are set to the current ``RMF''\footnote{nixtiref20170601v001.rmf} and 
``ARF''\footnote{nixtiaveonaxis20170601v002.arf} spectrum calibration 
files, respectively. Finally the ``BACKFILE'' keyword in the onpulse 
spectrum is set to the offpulse spectrum. This ensures that when 
the onpulse spectrum file is defined as the data in {\it{xspec}}, it 
automatically picks up the offpulse spectrum as the background spectrum, 
and also picks up the appropriate spectrum calibration files.

\begin{figure}[h]
\centering
\includegraphics[width=8.5cm]{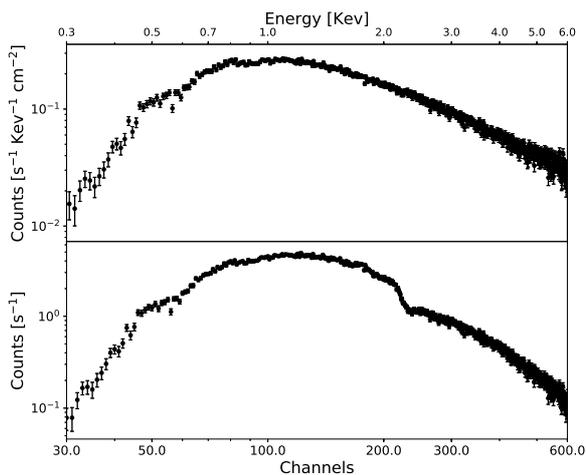}
\caption{
	Calibrated (top panel) and uncalibrated (bottom panel) spectrum
	of ObsID $1013010147$.
        }
\label{fig5}
\end{figure}

Finally, each onpulse spectrum is analyzed using {\it{xspec}}, which
is run in the ``PYTHON'' environment, using the ``from xspec import *''
command in PYTHON, which picks up the PYTHON libraries from the HEAsoft 
6.25 environment if the latter has been installed properly. Channels
below $30$ and above $600$ are ignored, the abscissa is set to the
energy mode (``Plot.xAxis = Kev''), and channel dependent effective 
area is set (``Plot.area = True''). Then the energy, count rate and its
error are printed out. The top panel of Fig.~\ref{fig5} displays the
calibrated spectrum for ObsID $1013010147$.

The uncalibrated spectrum (bottom panel of Fig.~\ref{fig5}) is obtained 
by removing the response file (``Spectrum.response = None'') and 
resetting the abscissa to channel mode (``Plot.xAxis = channel''). The 
uncalibrated spectrum can also be obtained by extracting the raw counts 
in each channel from the onpulse and offpulse spectra, using the {\it{ftlist}} 
tool, then dividing them by the respective live times, and then subtracting 
the two values; this is essentially what {\it{xspec}} also does, apart 
from calibration. As a consistency check, the integrated photon count 
rate of the uncalibrated spectrum (over all channels) should agree with 
the values in the top panel of Fig.~\ref{fig4}, after accounting for (a) 
number of detectors, and (b) the fact that onpulse photons in 
Fig.~\ref{fig5} arrive during only $80$\% of the period, while in 
Fig.~\ref{fig4} they are assumed (correctly) to arrive over the entire 
period, in order to estimate the average count rate.

\begin{figure}[h]
\centering
\includegraphics[width=8.5cm]{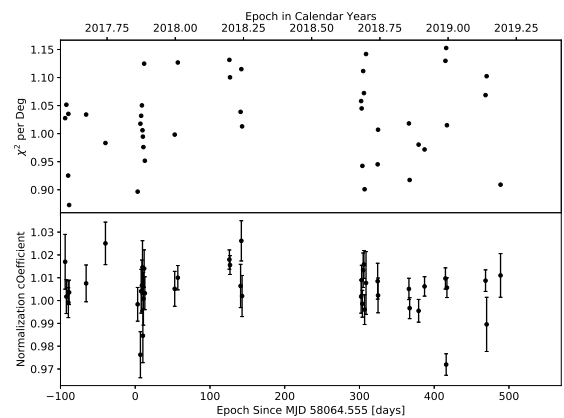}
\caption{
	Variation of spectral properties as a function of epoch.
	Top Panel: Normalized $\chi^2$ after subtracting the 
	reference calibrated spectrum from the calibrated 
	spectrum of the rest of the ObsIDs. Bottom panel: 
	Average Ratio after dividing the calibrated spectrum 
	of each ObsID with the reference calibrated spectrum.
        }
\label{fig6}
\end{figure}

Fig.~\ref{fig6} shows the result of comparing the calibrated spectrum of
each ObsID with that of the reference ObsID ($1013010147$). The top
panel of Fig.~\ref{fig6} shows the $\chi^2$ per degree of freedom
after subtracting the calibrated spectrum of each ObsID from the
calibrated spectrum of the reference ObsID ($1013010147$). The values are
very close to $1.0$, with negligible error on them since the number
of degrees of freedom is $570$. The mean preglitch and postglitch $\chi^2$ 
values are $0.99$ and $1.03$, with standard deviations $0.06$ and $0.07$
respectively. The bottom panel conducts the same 
comparison in a different way. The calibrated spectrum of each ObsID is
divided by the calibrated spectrum of the reference ObsID ($1013010147$),
channel by channel, and the average ratio across the spectrum is plotted.
The values lie close to $1.0$; the mean preglitch and postglitch ratios are 
$1.01$ and $1.00$, with standard deviations $0.01$ and $0.01$ respectively.
In both panels of Fig.~\ref{fig6}, the scale of variation is similar before
and after the glitch.

Fig.~\ref{fig6} has also been produced using the uncalibrated spectra, and 
the results are almost identical.
This section therefore concludes that there is no difference in the
soft X-ray spectrum of the Crab pulsar before and after the glitch.

\section{Giant pulses}

The Crab pulsar emits at radio frequencies what are known as giant pulses;
their amplitude is orders of magnitude larger than that of the mean radio 
pulse.  These occur only within the main pulse and inter pulse, which at 
radio frequencies are very narrow windows in phase. The radio peaks are 
aligned with the corresponding peaks at X-rays \citep{Lundgren1995, 
Sallmen1999, Hankins2000, Cordes2004, Karuppusamy2010}. Several attempts 
to detect giant pulses in X-rays and $\gamma$-rays have proved futile 
\citep{Bilous2012, Mickaliger2012, Ahronian2018}.

\begin{figure}[h]
\centering
\includegraphics[width=8.5cm]{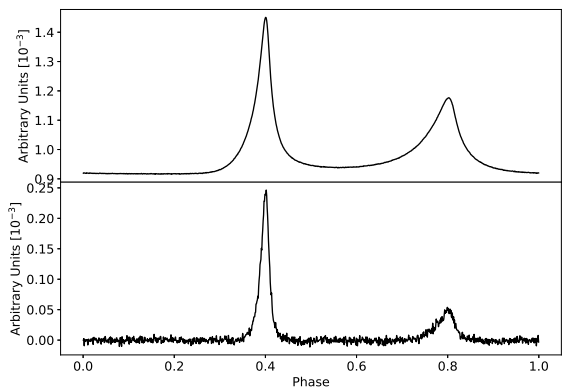}
\caption{
	Test for giant pulses.
	Top Panel: Integrated profile of the Crab pulsar in soft X-rays,
	using $289.3$ Ksec of data ($3181769922$ photons from $51$ ObsIDs)
	in$1024$ phase bins. Bottom panel: Difference between the IP of 
	the top panel, and the IP obtained by including only those photons 
	that differ in arrival times by less than $1$ $\mu$s from each other, after proper 
	scaling, as explained in the text.
        }
\label{fig7}
\end{figure}

Fig.~\ref{fig7} top panel shows the IP of the Crab pulsar (in arbitrary units)
using most of the data acquired by NICER, after aligning the IP of each ObsID
with that of the reference ObsID $1013010147$, using cross-correlation. Visual 
inspection of the IP, particularly in the phase range $0.2$ centered at the 
main and inter pulses (see Fig.~$2$ of \cite{Mickaliger2012}; see also Section $7$ of 
\cite{Cordes2004}), revealed no indication of an excess emission. One expects
this excess emission to be confined to a few phase bins that represent the 
domain of arrival of giant pulses.

To investigate this issue more quantitatively, an IP was formed from the data
using only those photons that differed in arrival times by less than one $\mu$s 
from each other. If the giant pulse phenomenon at radio wavelengths implies more 
simultaneous photons at X-rays also, then the second IP should show an excess at 
the phases where giant pulses are expected. This IP contained a factor of 
$\approx 24$ less photons than that in the top panel of Fig.~\ref{fig7}. After 
proper scaling of both IPs (total area under the IPs should be $1.0$) and 
subtracting, and after introducing a scaling constant, the result obtained is 
shown in the bottom panel of Fig.~\ref{fig7}. There is no excess of photons that 
is confined to very few channels; the excess is spread over the phase ranges of 
the main and inter pulses, which is expected naturally because the probability 
of arrival of closely spaced photons increases with increasing photon count.
The photons in the bottom panel of Fig.~\ref{fig7} arrive on 
different detectors, since on any single detector the time interval between 
any two photons can not be less than the dead time ($15$ or $22$ $\mu$s).

\section{Timing noise}

\begin{figure}[h]
\centering
\includegraphics[width=8.5cm]{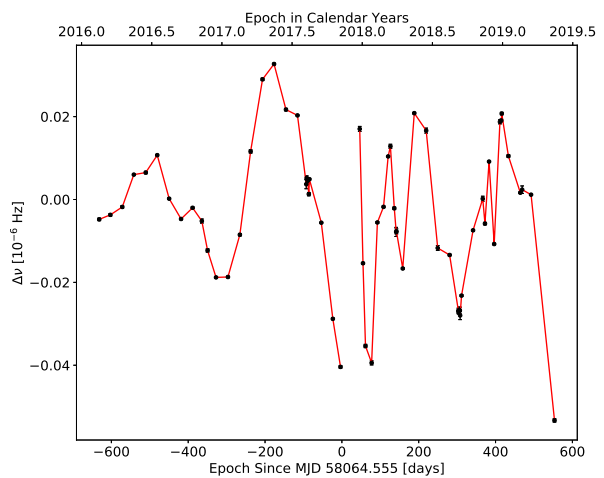}
\caption{
	Timing noise of the Crab pulsar for about $3.5$ years centered on 
	the glitch.
        }
\label{fig8}
\end{figure}

Timing noise is the slow and irregular variation of the rotation frequency 
($\nu$) of a pulsar \citep{Lyne1993, Hobbs2010, Shannon2010, Melatos2014}, 
unlike the abrupt change in $\nu$ at a glitch, and is as yet an unexplained
phenomenon. Here it is demonstrated that the largest glitch in the Crab 
pulsar may have modified its timing noise.

\begin{table}
\begin{center}
\caption{Glitch model, applied to the postglitch data in the top panel of 
	Fig.~\ref{fig2}, for data from $20$ to $554$ days after the glitch. 
	The subscript ``p'' refers to permanent changes due to the glitch, 
	while the subscript ``n'' refers to exponentially decaying values.
\label{tbl2}
}
\begin{tabular}{|l|c|}
\hline
Parameter  & Value \\
\hline
Glitch Epoch (MJD) & $58064.555$ \\
\hline
$\Delta \nu_p$ ($10^{-6}$ Hz) & 4.4(1) \\
\hline
$\Delta \dot \nu_p$ ($10^{-12}$ Hz s$^{-1}$) & $-0.456(7)$ \\
\hline
$\Delta \ddot \nu_p$ ($10^{-20}$ Hz s$^{-2}$) & $0.18(2)$ \\
\hline
$\Delta \nu_n$ ($10^{-6}$ Hz) & $10.62(8)$ \\
\hline
$\tau$ (days) & $56(1)$ \\
\hline
\hline
\end{tabular}
\end{center}
\end{table}

Fig.~\ref{fig8} shows the timing noise of the Crab pulsar for $1200$ days.  
The preglitch data is identical to the preglitch data in the top panel 
of Fig.~\ref{fig2}, where only $100$ days of this data are displayed. The 
postglitch data in Fig.~\ref{fig8} are the residuals obtained after 
fitting Eq.~$1$ of \cite{Vivekanand2015} (or Eq.~$1$ of \cite{Zhang2018})
to the postglitch data in the top panel of Fig.~\ref{fig2}, using the 
glitch parameters given in Table~\ref{tbl2}. The fit is done to data from 
$20$ to $554$ days after the glitch epoch. Data from the glitch epoch to 
day $20$ varies quite differently, as is evident in the top panel of 
Fig.~\ref{fig2}, but is more clearly apparent in panel C of Fig.~$1$ 
of \cite{Shaw2018}. Attempt was made to fit this data in two segments -- 
data from glitch epoch to day five, and data from day five to day $20$. The
resulting residuals were very noisy, and so have been omitted from
Fig.~\ref{fig8}.

The above fit excludes the latest reasonably sized glitch on MJD $58687.59$
but includes the two much smaller glitches on MJDs $58237.357$ and $58470.939$,
which fall on days $172.8$ and $406.4$ in Fig.~\ref{fig8}. Their
magnitudes are $0.008$ and $0.005$ fraction of the largest glitch.

The slow and stochastic changes in the $\nu$ of the Crab pulsar with 
respect to its secular variation are evident, with time scales of 
$\approx$ a few hundred days before the glitch,  and shorter time 
scales after the glitch, even if one restricted oneself to the range $20$
to $172$ days after the glitch. The quasi sinusoidal variation before the 
glitch, with increasing amplitude, is absent after the glitch. However,
since the data of two small glitches have been included in this analysis,
the postglitch results of this section can only be tentative.

The difference in preglitch and postglitch $\Delta \nu$ in Fig.~\ref{fig8} 
can not be attributed to to any difference in analysis method.  Firstly, 
there is no bias in actual stride lengths between preglitch and postglitch 
analysis. Secondly, the X-ray data in Fig 8 (points having larger error 
bars) fall very close to their corresponding radio points (having no
visible error bars at all), when the two observational epochs nearly 
coincide, say at epochs -100 and 500 days along abscissa.

\section{Discussion}

The summary of this work is the following.
Firstly, the $\nu$ and $\dot \nu$ of the Crab pulsar vary similarly at the 
radio and soft X-ray energies, during the glitch of $2017$ Nov $8$.
Secondly, the following properties of the Crab pulsar remain essentially
constant before and after the glitch: (a) the total onpulse X-ray flux, 
(b) the flux of the two components of its integrated profile, (c) the 
widths and peaks of these two components, and (d) the soft X-ray spectrum. 
Thirdly, there is no evidence for giant pulses at X-ray energies.
Finally, the timing noise appears to change after the glitch.
The following discussion focuses on three aspects of the radio and X-ray 
emission of the Crab pulsar, and on the requirement for X-ray timing
observations of rotation powered pulsars.

\subsection{Overall emission}

Even the strongest glitch in the Crab pulsar (comparable to those occurring
in the Vela pulsar) does not change several emission parameters discussed
above, both at radio as well as X-ray energies. This is consistent with the
belief that both emissions occur very close to each other in the pulsar's
magnetosphere. This has been independently demonstrated by \cite{Rots2004} 
and \cite{Bilous2012}, who show that the separation in phase between the 
X-ray and radio peaks is $0.01$, the radio peak arriving later, so the
two emission regions are separated by $\approx 0.01 / 29.63 * 3 \times 
10^5 \approx 100$ kilo meters (km), which is a small fraction of the
light cylinder radius in the Crab pulsar ($\approx 1600$ km).

The results of this work are also consistent with the belief that both
emissions occur far away from the surface of the neutron star, say in
the outer gaps \citep{Romani2010}, since any effect of the glitch is
more likely to occur closer to the surface of the neutron star. However,
this argument needs to be considered with some caution. Only two
pulsars (out of about two thousand)  have shown glitch associated 
radiative changes, and as argued by \cite{Shaw2018}, it is not clear
why a fractional increase of $\approx 10^{-6}$ in rotation frequency
at the surface of the neutron star should induce changes in the
pulsar magnetosphere that could affect the pulsar emission.

Since the X-ray spectrum is also unchanged on account of the glitch,
the electric fields that accelerate electrons and positrons in the 
outer gaps, their energy distribution, and the magnetic field in the 
magnetosphere where they emit X-rays, all three remain the same 
before and after the glitch. However, the charged particles begin
their life at the surface of the neutron star, in the polar cap
\citep{Romani2010}, so if the glitch is accompanied by a
re-arrangement of the surface (say a star quake), some of the above 
three parameters are likely to get affected. Even then, we are
unlikely to observe the consequences if the intense gravity at the
surface of the neutron star re-adjusts the surface on time scales
of, say, $1000$ sec, which is the time required to obtain a 
significant integrated profile.

\subsection{Giant pulses}

The phenomenon of radio giant pulses does not yet have an explanation; see
\cite{Lundgren1995, Bilous2012, Mickaliger2012} and references therein 
for some possibilities. The absence of simultaneous giant pulses at
X-rays and $\gamma$-rays could imply  that coherent bunching of 
emitting particles might occur on length scales that are much smaller
than radio wavelengths, but much larger than X-ray wavelengths.
Before proceeding further it should be mentioned that some evidence
for giant pulses has been reported at optical wavelengths 
\citep{Shearer2003}. Here it is argued that there may be instrumental 
selection effects in observing giant pulses at X-rays.

Let us consider the various possibilities at X-rays that can accompany
a radio giant pulse, if the two are indeed correlated. First, there
can be multiple X-ray photons emitted simultaneously; Fig.~\ref{fig7}
tries to explore this possibility. However, the veto sections of X-ray 
detectors may discard such photons, because they might be misconstrued 
for high energy particles, depending upon how many photons there are
and what their energy is and what kind of detectors are used, unless 
there is reasonable time separation between the simultaneous photons.
Moreover, this time separation must be larger than the dead time of 
the detector. The design of NICER drastically reduces the dead time 
problem -- there are several detectors to receive the simultaneous 
photons. However, if the giant pulses at X-rays are emitted by a 
coherent mechanism (such as in a laser), and they all happen to fall on
one single detector of NICER, then one might have to once again face
the veto and dead time aspects of the X-ray detector.

Next, instead of several simultaneous X-ray photons, maybe only a single
higher energy photon is emitted; this speculation is not unjustified 
since we do not know the exact mechanism of giant pulse emission. Then, 
once again depending upon the actual situation, the photon may either get 
vetoed, or it falls in a higher energy channel. In this case, one
would be looking for excess photons in a region of the spectrum that
naturally has very few photons, due to the spectral index of the
Crab pulsar at X-ray energies. An inspection of the spectrum for the
combined data in Fig.~\ref{fig7} revealed no such excess. If the
energy of such a photon is larger than $12$ Kev, then NICER would
anyway not detect it.

Finally, a radio giant pulse may be accompanied by just a single 
average energy X-ray photon. In this best scenario situation, the 
better way of looking for giant pulses using NICER may be to observe 
simultaneously at radio and X-rays and look for photons that are 
correlated at both wavelengths, as was done by \cite{Lundgren1995, 
Bilous2012, Ahronian2018}.

\subsection{Pre glitch timing noise}

The preglitch timing noise in Fig.~\ref{fig8} looks like an oscillating
signal with increasing amplitude. This requires two ingredients for
explanation -- an oscillator, and a positive feed back system.

\cite{Ruderman1970} proposed that the quasi periodic timing in the Crab
pulsar could be due to Tkachenko oscillations, in which the vortices
remain parallel to the rotation axis, but their density (and therefore
the angular momentum of the fluid) are redistributed periodically.
Equation~$7$ of \cite{Ruderman1970} shows that the periods of these
oscillations are $\approx 120$ days, which is similar in order of
magnitude to the $\approx 200$ days quasi periodicity seen before the 
glitch. 

Other modes of oscillation, such as the Kelvin modes or the r-modes, 
have periods much much smaller than the  above values \citep{Ruderman1970, 
Haskell2015, Haskell2018}. Therefore, for the purpose of this discussion 
Tkachenko oscillations are assumed to be operative before the glitch; 
however, see \cite{Haskell2011} for the effect of vortex pinning and fluid 
compressibility, and maybe even entrainment, on the viability of these 
oscillations.

Next one needs a positive feed back system, something that pumps energy
into the oscillator at the correct phase to enhance the amplitude. It is
speculated here that this could probably be due to some vortices that are 
pinned strongly, and therefore are not moving, but are still executing 
the Tkachenko oscillations, albeit weakly, due to the ``knock on'' effect 
of vortices that are in the process of creating an ``avalanche''; see 
\cite{Haskell2015, Haskell2018} for details. These vortices might strain 
the crustal lattice, and could create small cracks in it, which would 
create heat that is pumped into the super fluid system. This periodic 
thermal pulse could create a periodic bias in the potential in which the 
vortices are ``creeping'' \citep{Alpar1984}. Depending upon the 
actual numbers and one's imagination, one could believe in the 
possibility of a positive feed back.

Two points need to be clarified regarding the above speculation.
First, the vortices providing the positive feed back are not the
ones that are annihilating themselves while depositing angular 
momentum into the crust. They belong to a different domain in 
the neutron star, they are pinned to the crust, and they merely 
sense the Tkachenko oscillations, 
by virtue of being connected to the super fluid. The high
thermal conductivity of the neutron star might ensure that this
heat percolates into the neighboring domains where the vortices
are free to move. Second, the scale of this activity is orders of 
magnitude smaller than the scale of the glitch of $2017$ Nov $8$ 
-- see the difference in the scale of the ordinates in Figures 
\ref{fig2} and \ref{fig8}.
Finally, it is interesting that, while the giant glitch of $2017$ 
Nov $8$ in the Crab pulsar does not fall within the epoch 
predicted by \cite{Vivekanand2017}, one smaller and one relatively
larger glitch 
(MJDs $58470.939$ and $58687.59$)\footnote{http://www.jb.man.ac.uk/pulsar/glitches/gTable.html}
lie within the uncertainty of prediction.

\subsection{Importance of X-ray timing observations}

Most rotation powered pulsars are easier to observe at radio frequencies than at X-ray energies,
for reasons of sensitivity, cost and convenience. However there are several scientific reasons
for observing them at X-rays also.
Firstly, in some pulsars like the Crab, glitches are believed to be triggered by cracking of
the crust of the neutron star, which can lead to bursts of X-ray emission \citep{Ruderman1991}.
Therefore X-ray study of pulsar glitches is important to understand their trigger mechanism.
Next, the much wider main and inter pulse of the Crab pulsar at X-rays, relative to their
radio counterparts, implies that study of pulse morphology variation at X-rays might sense
the properties of a much larger volume of the pulsar magnetosphere, than such studies at radio
frequencies.
Finally, optical and X-ray studies of giant pulses in pulsars might reveal the size of 
the basic unit of emission, if the phenomenon is due to decreasing coherence of the emission 
region at increasing frequencies of observation.

\begin{acknowledgements}
I thank Marco Antonelli and Brynmor Haskell for useful discussion regarding 
the role of vortices in the neutron star super fluid. I thank the referee for 
useful discussion.
\end{acknowledgements}

\begin{appendix} 

\section{Details of stride fitting}

\begin{figure}[h]
\centering
\includegraphics[width=8.5cm]{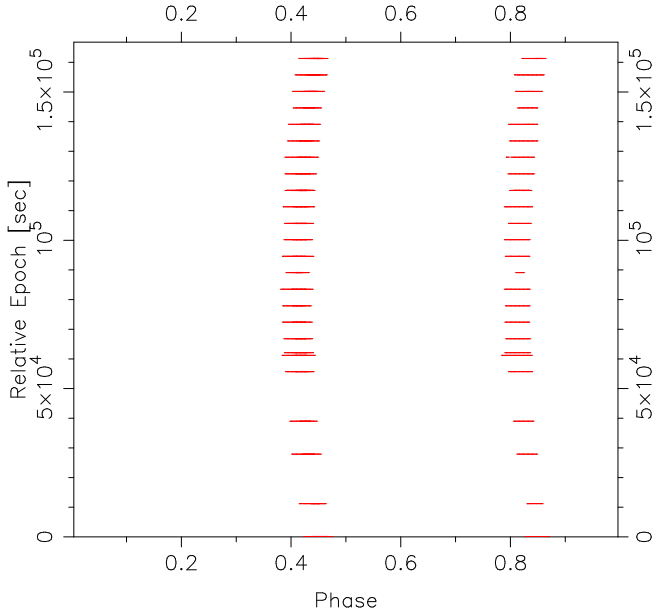}
\caption{
        Contour plot of the combined data of the ObsIDs $1013010116$, 
	$1013010117$, and $1013010118$, after alignment using initial 
	$\nu$ and $\dot \nu$, whose estimation is explained in Appendix 
	A. The total data duration is about $170$ kilo seconds, while
	the abscissa is the phase of the photon event. Contrast has been
	adjusted to highlight the main pulse of the Crab pulsar, so the
	rest of the data may not be visible, but for parts of the inter 
	pulse data.
        }
\label{figa1}
\end{figure}

The stride fitting procedure is applied to the data of usually three, but 
sometimes two, consecutive ObsIDs. First, one has to estimate the initial
$\nu$ and $\dot \nu$ valid for all three (or two) ObsIds, starting from the 
intermediate $\nu$ and $\dot \nu$ of each ObsID that are derived in Section 
$2$. It was found that for most of the data the initial $\dot \nu = \dot 
\nu_0$ (Table~\ref{tbl1}) was a good starting point. This value had to be
modified only for data close in epoch to the epoch of minimum $\Delta \dot \nu$ in the 
lower panel of Fig.~\ref{fig2}. Using this value of initial $\dot \nu$, 
that value of initial $\nu$ was obtained, which best aligned the data of 
all three (or two) ObsIDs. The cross correlation of the IPs
of the first and second halves of the combined data was used as a measure 
of the alignment. These initial $\nu$ and $\dot \nu$ are referenced to the
epoch at the middle of the combined data. Fig.~\ref{figa1} is a contour 
plot of the combined data of the ObsIDs $1013010116$, $1013010117$, and 
$1013010118$. The initial $\nu = 29.6364286361$ Hz and the initial $\dot 
\nu = -3.606042 \times 10^{-10}$ Hz/sec; the latter is different from the 
$\dot \nu_0$ of Table~\ref{tbl1} to enhance the phase variation, for 
better viewing.

\begin{figure}[h]
\centering
\includegraphics[width=8.5cm]{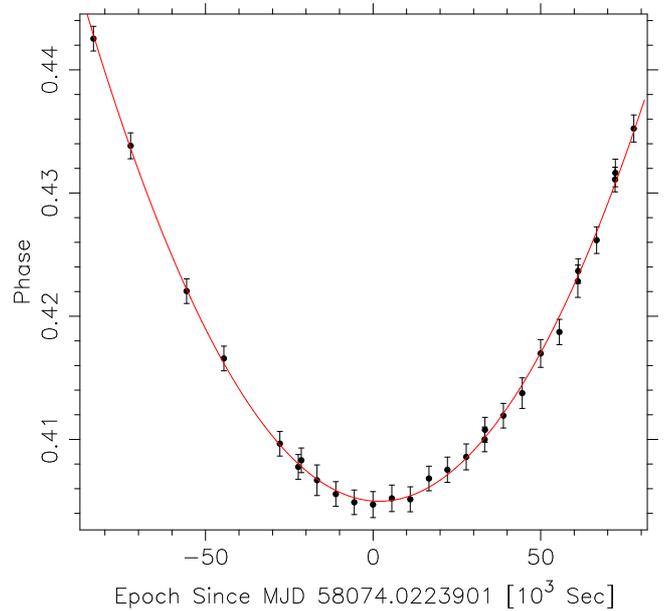}
\caption{
	Phase of the main pulse of the IP of $100$ sec
	duration, of the data in Fig.~\ref{figa1}.
        }
\label{figa2}
\end{figure}

Next, IPs are formed from data of $100$ sec duration, 
using the initial $\nu$ and $\dot \nu$ , and the phase of the main pulse 
is obtained by fitting a Gaussian to the peak (see \cite{Vivekanand2015} 
for details); this phase is plotted in Fig.~\ref{figa2} as a function of 
epoch, relative to the mid epoch of the combined data. The phase data shows the 
characteristic quadratic behavior expected for deviations of $\nu$ and 
$\dot \nu$ from their initial values. A second order polynomial is fit to
this phase data, to obtain the final $\nu$ and $\dot \nu$. It is essentially 
to such data that \cite{Shaw2018} and \cite{Zhang2018} have fit a second 
order polynomial, to obtain their $\nu$ and $\dot \nu$ as a function of epoch.

In Fig.~\ref{figa2}, a correction to $\nu$ will shift the position of 
minimum phase away from the origin. It is clear that this correction is 
quite small; it turn out to be $-0.019(2)$ $\mu$Hz. The correction to
$\dot \nu$ is $0.104(1) \times 10^{-10}$ Hz/sec. It is important to get
the initial values of $\nu$ and $\dot \nu$ as close as possible to the final
values. Otherwise, the quadratic curve would be shifted away from the origin
and there is a chance of the higher phase values becoming larger than 1.0,
in which case they are folded over to very small values in the next 
cycle.

\end{appendix}
\end{document}